\documentclass[10pt,twocolumn]{article}

\usepackage{txfonts}
\usepackage{natbib}
\usepackage{graphicx}
\usepackage{journals}
\usepackage[colorlinks,citecolor=blue,urlcolor=blue,filecolor=blue,linkcolor=blue]{hyperref}
\usepackage{xcolor}

\newcommand{\orcidlogo}{\includegraphics[height=\fontcharht\font`A]{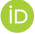}}
\newcommand{\orcid}[1]{\href{https://orcid.org/#1}{\orcidlogo}}

\graphicspath{{./}{figures/}}
\def\arcmin{\hbox{$^\prime$}}

\begin{document}

\title{Simultaneous broadband radio and optical emission of meteor trains imaged by LOFAR / AARTFAAC and CAMS}

\author{Tammo Jan Dijkema$^1$\orcid{0000-0001-7551-4493}, Cees Bassa$^1$\orcid{0000-0002-1429-9010}, Mark Kuiack$^2$\orcid{0000-0002-8899-434X}, Peter Jenniskens$^3$\orcid{0000-0003-4735-225X},\\ Carl Johannink$^3$, Felix Bettonvil$^4$, Ralph Wijers$^2$\orcid{0000-0002-3101-1808}, Richard Fallows$^1$\orcid{0000-0002-5186-8040}}

\date{%
$^1$ASTRON, Netherlands Institute for Radio Astronomy\\
$^2$Anton Pannekoek Institute, University of Amsterdam\\
$^3$SETI Institute\\
$^4$Leiden University / NOVA\\[2ex]
November 18, 2021}

\maketitle

\begin{abstract}
  We report on simultaneous 30 - 60 MHz LOFAR / AARTFAAC12 radio observations 
  and CAMS low-light video observations of +4 to -10 magnitude meteors at the peak 
  of the Perseid meteor shower on August 12/13, 2020. 204 meteor trains were imaged in 
  both the radio and optical domain. Aside from scattered artificial radio
  sources, we identify broadband radio emission from many persistent trains,
  one of which lingered for up to 6 minutes. Unexpectedly, 
  fewer broadband radio meteor trains were recorded when the experiment was
  repeated during the 2020 Geminids and 2021 Quadrantids.
  Intrinsic broadband radio emission was
  reported earlier by the Long Wavelength Array, but for much brighter meteors and
  observed with lower spatial resolution. The new results offer insight into
  the unknown radio emission mechanism.
\end{abstract}

\section{Introduction}
Meteors are well-known to reflect artificial radio emission in 
forward and backward scattering. Combined optical and radio meteor 
scatter observation of meteors go back to \citep{1947MNRAS.107..155P}, 
showing that meteor trains are overdense for most visible meteors. Forward
meteor scatter, where transmitter and receiver are not at the same location,
can be used to study meteors as is done in, e.g., the 
Belgian RAdio Meteor Stations (BRAMS) network \citep{2011MSSB}. 

Less well established is the detection of intrinsic natural radio emission 
from the meteor or meteor train itself. Early efforts to detect intrinsic
emisison relied on temporal coincidences (e.g., \citep{2000EMP82-83}).
The first more substantive reports based on radio imaging
came from all-sky imaging at low radio frequencies ($\sim$40\,MHz) with the Long
Wavelength Array, detecting intrinsic non-thermal radio emission
from fireballs 
\citep{2014ApJ...788L..26O,2016AGUFM.P33E..06O,2016JGRA..121.6808O}. The
emission persisted well after the meteor itself had faded. The
emission mechanism of these meteor radio afterglows is not fully
understood, and so far has not been independently confirmed with other
instruments. A survey with the Murchison Widefield Array
\citep{2018MNRAS.477.5167Z} did not show this intrinsic
emission at higher 72--103\,MHz frequencies.

Several mechanisms for broadband radio emission from meteors have been suggested in the
literature, namely reflected broadband terrestrial radio emission, 
intrinsic emission from chemically produced suprathermal electrons
\citep{2020JGRA..12528053O}, Langmuir waves
\citep{2015JGRA..120.9916O}, free-free emission
\citep{2018Ge&Ae..58..693F} and transition radiation
\citep{2020JGRA..12528053O}. Also, bright celestial radio emission shines on the meteor
trains and that radiation may be scattered, as suggested in \cite{2015JGRA..120.9916O}.

Here we report on low-frequency all-sky observations with the LOFAR
(``Low Frequency Array'') radio telescope \citep{2013A&A...556A...2V}
during the 2020 Perseid meteor shower \citep{2006MeteorShowers}. 
The radio data are complemented with
simultaneous, both temporally and spatially, optical video observations
from the CAMS BeNeLux network \citep{2011Icar..216...40J} in an effort to
study the altitude dependence of the proposed intrinsic radio emission.

\section{Observations}
The LOFAR radio telescope consists of thousands of dipole antennas
grouped into stations spread over Europe, with a dense core of
stations located in the North of the Netherlands. Each LOFAR station
has low-band antennas (LBAs) which can observe from 10 to 90\,MHz and
high-band antennas (HBAs) operating from 110 to 250\,MHz. During
regular LOFAR LBA observations radio signals are combined (beamformed)
hierarchically, by first combining radio signals from dipole antennas
in a station, and then combining or correlating the combined radio
signals from stations to increase sensitivity at the expense of
field-of-view.

\begin{figure}
  \includegraphics[width=\linewidth]{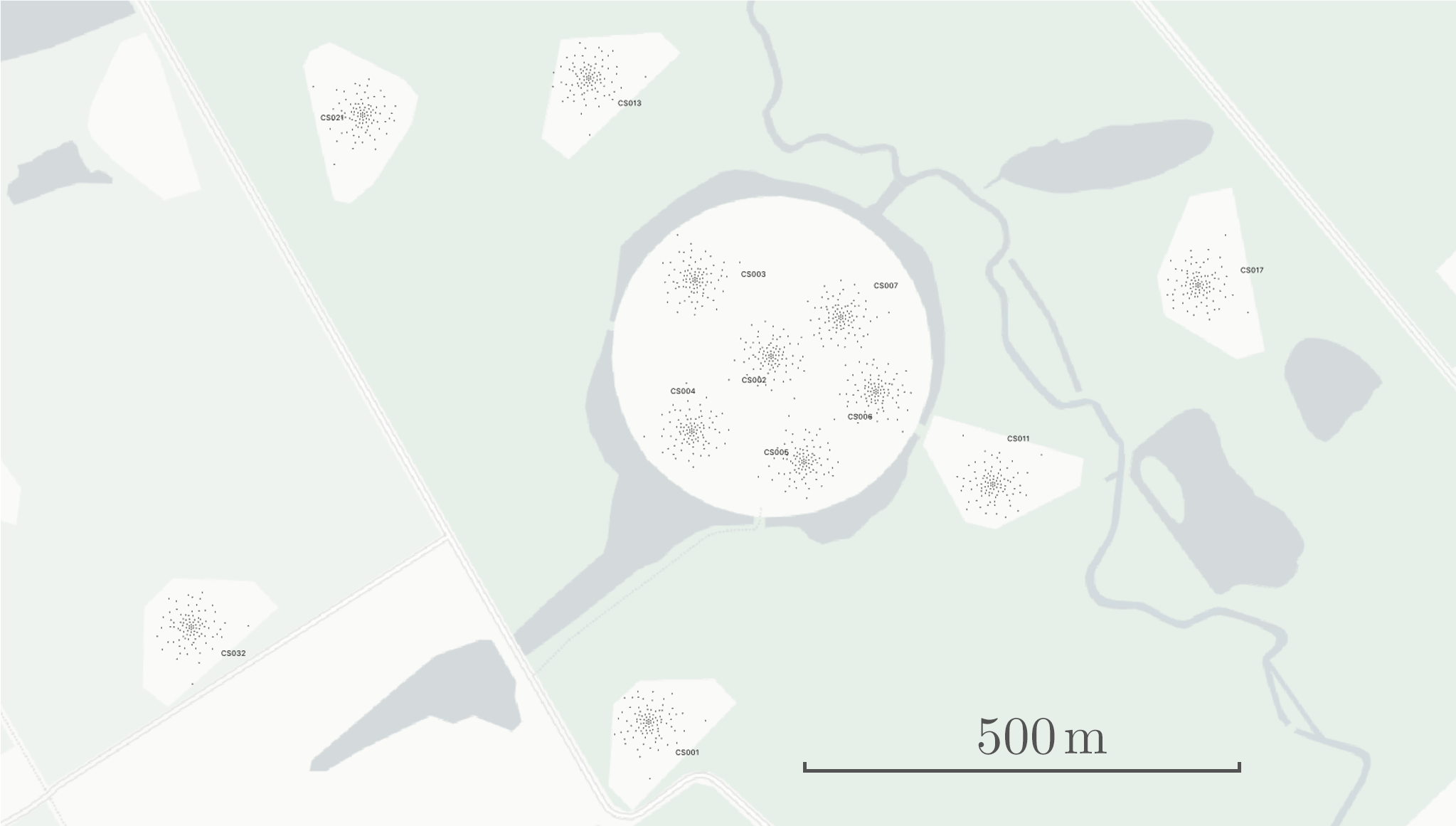}
  \includegraphics[width=\linewidth]{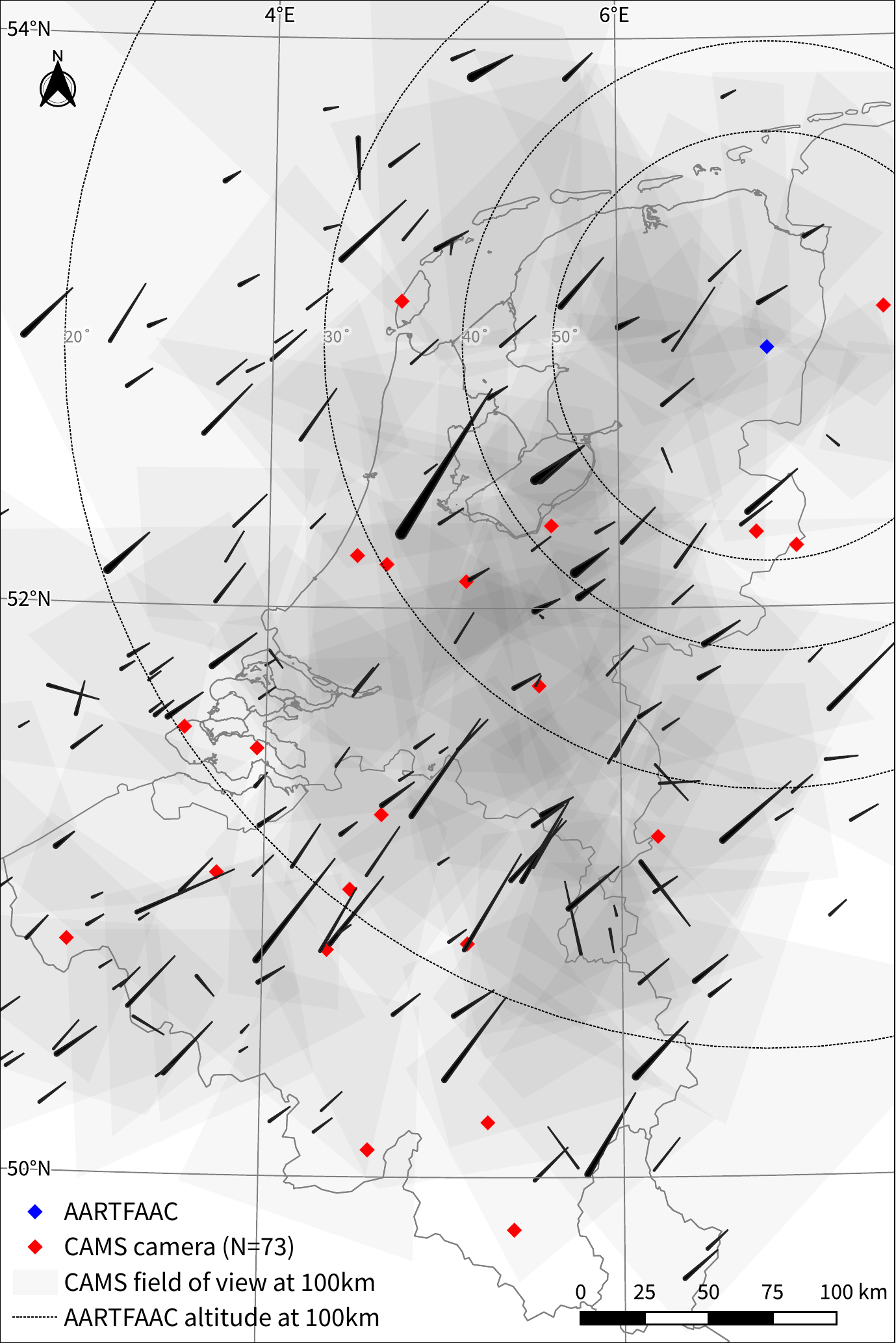}
  \label{fig:meteormap}
  \caption{Top: map of the LOFAR core with the location of all 512 dipole
    antennas used for AARTFAAC. Bottom: map with CAMS camera locations (red),
    the location of the LOFAR core (blue) and co-observed meteors.
  }
  \label{fig:camsmap}
\end{figure}

The AARTFAAC piggyback instrument
\citep{2016JAI.....541008P,2019MNRAS.482.2502K,2021MNRAS.505.2966K}
provides a special observing mode whereby LOFAR can create all-sky
images of the radio sky. It does so by correlating the radio signals
from all 576 LBA dipoles from the inner 12 LOFAR stations located in
the dense core (with a 1.2\,km diameter) surrounding the ''super terp'' 
in Drenthe, the Netherlands, see Fig.\,\ref{fig:camsmap}.

LOFAR/AARTFAAC observations were obtained during the Perseid meteor
shower (2020 August 12-13) for 64 frequency channels of 48.8\,kHz
bandwidth (grouped into 16 subbands of 4 channels each), spread over
radio frequencies between 30 to 60\,MHz. These observations were used
to create all-sky images with $3\arcmin$ spatial resolution and 2\,s
time resolution.

\begin{figure}[!ht]
  \centering
  \parbox{.72\linewidth}{
  \includegraphics[width=\hsize]{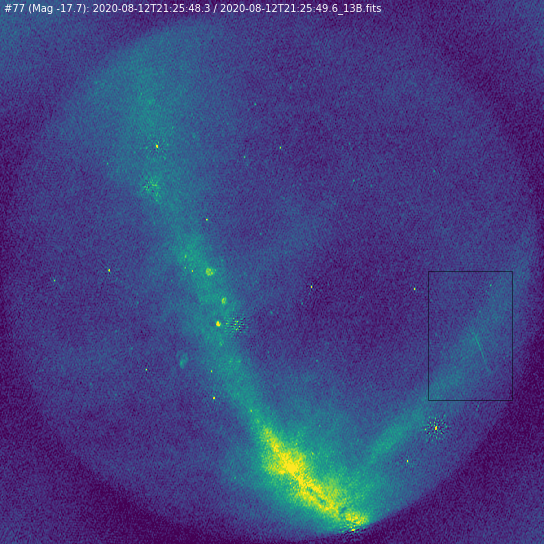}}
  \parbox{.243\linewidth}{\includegraphics[width=\hsize]{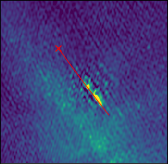}
  \includegraphics[width=\hsize]{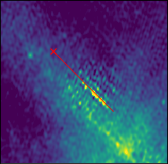}
  \includegraphics[width=\hsize]{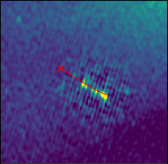}
  }
  \parbox{.72\linewidth}{
  \includegraphics[width=\hsize]{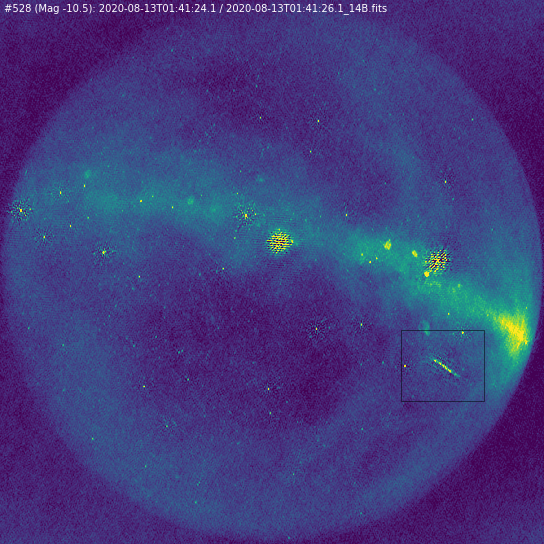}}
  \parbox{.243\linewidth}{\includegraphics[width=\hsize]{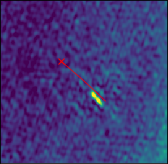}
  \includegraphics[width=\hsize]{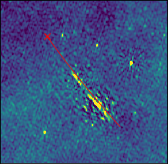}
  \includegraphics[width=\hsize]{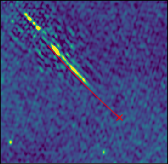}
  }
  \caption{AARTFAAC image (integrated over all observing bands) of the
    meteors at 2020-08-12 21:25:48 UTC (top left) and 2020-08-13 01:41:24 UTC (bottom left). The large-scale diffuse emission is the Galactic plane, the bottom left image shows some residuals of the subtracted sources Cassiopeia~A and Cygnus~A. The right column shows zoomed in meteors, from top to bottom corresponding to numbers 144, 235, 249, 317, 333, and 704 in the CAMS data set. In red, the trajectory as computed from the CAMS optical observations is
    overlaid.}
    \label{fig:camsmatch77}
\end{figure}

All-sky radio images were created for each of the individual AARTFAAC
subbands (a set of 4 channels). This analysis used a newer version of the
pipeline described in \citet{2021arXiv210315160S}. As part of the
pipeline, the brightest radio sources Cassiopeia\,A and Cygnus\,A were
subtracted, to minimize imaging artefacts across the all-sky
images. Typically, in subbands containing strong narrow-band
terrestrial radio emission reflected by meteor ionization trails, such
as emission from the BRAMS meteor forward scatter network transmitter in Belgium near 50\,MHz
\citep{2016EGUGA..1811624L}, the imaging failed, and images from these
subbands were rejected. Images from the remaining subbands were
averaged in radio frequency across subbands. This resulted in 10713
all-sky radio images at a 2\,s cadence, corresponding to 5\,h of
data. Example all-sky images obtained with AARTFAAC are shown in
Fig.\,\ref{fig:camsmatch77}. A full time-lapse of these images is available as \cite{zenodo_mp4_perseids}.

The CAMS BeNeLux low-light video network is part of the global 
(''Cameras for Allsky Meteor Surveillance'') network and uses close to
100 low-light video cameras spread over the BeNeLux to triangulate optical 
meteors +4 and brighter using methods 
described in \cite{2011Icar..216...40J}. Weather was clear during 
the night of 2020 August 12-13 and the trajectory and orbits of 720 
meteors were measured \citep{2020eMetN...5..400R}.

Figure\,\ref{fig:camsmatch77} shows a subset of these meteor
trajectories overlaid on the all-sky radio images from the perspective
of the ''super terp''. All all-sky radio
images within a few seconds of CAMS detections were inspected manually
for radio emission coincident with the reconstructed meteor
trajectory. This resulted in 204 meteors where radio emission was
coincident with a CAMS detection. Of these meteors, 59 had a
discernible trail in the radio images, the others showed up as
point-like in the radio images. There were also dozens of radio
meteors without a counterpart in the CAMS data, mostly due to missing
sky coverage of CAMS in Germany at the time, see Fig.\,\ref{fig:camsmap}.

For each radio meteor coincident with a CAMS detection we have
determined the begin and end point of the radio trail in the all-sky
radio images and the duration for which the meteor was visible in the
radio images.

\section{Results}
\begin{figure}[!ht]
  \centering
  \includegraphics[width=\linewidth]{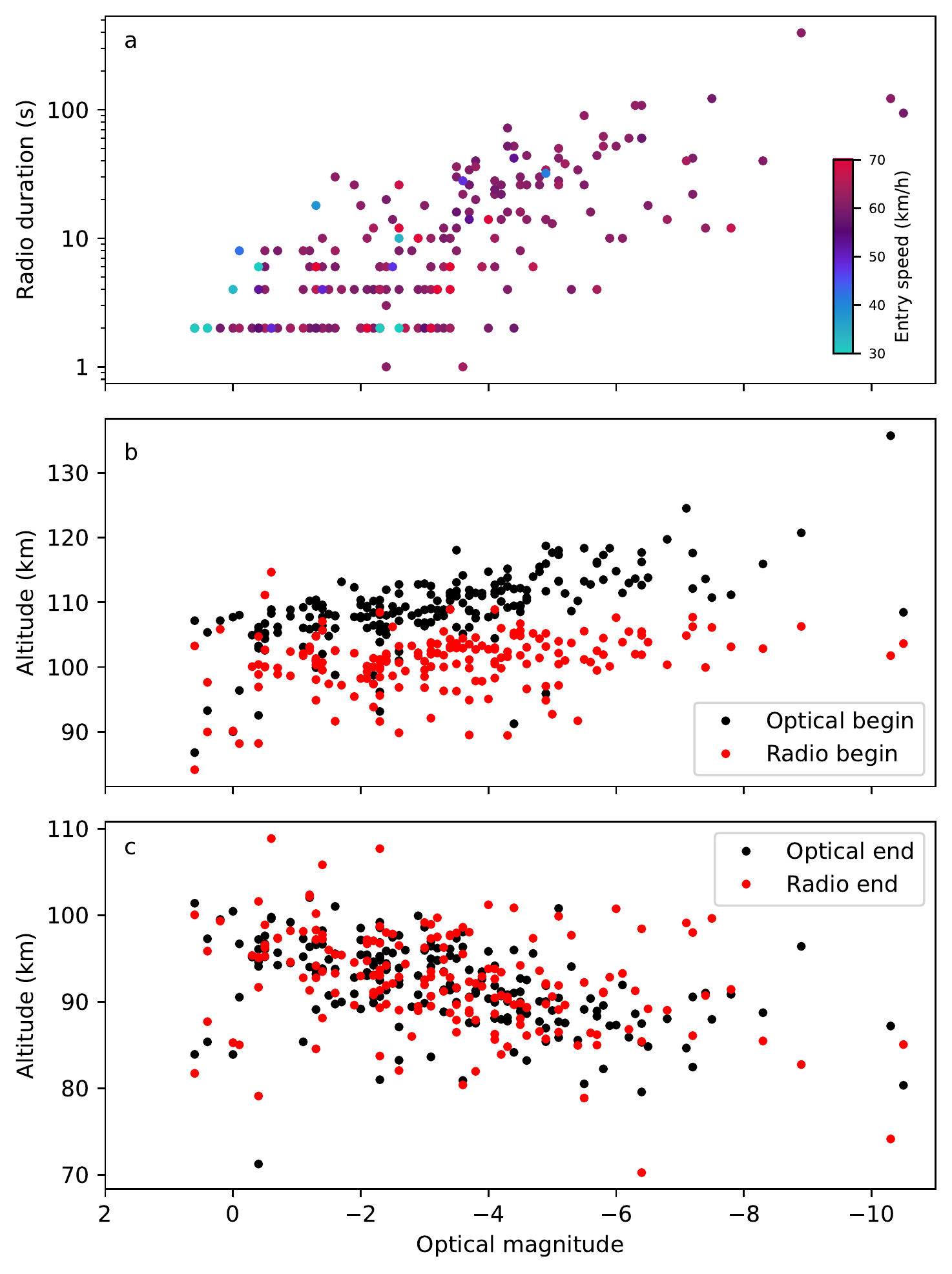}
  \caption{The dependency of the radio visibility and the altitude of
    the beginning and end of the meteor trail with optical
    brightness and entry speed. Note that the CAMS-derived peak optical brightness is less
    reliable below -5 magnitude due to video blooming.}
  \label{fig:magnitude}
\end{figure}

The coincidence of optical and radio trails demonstrates that meteors
are a source of radio emission, be it scattering or intrinsic emission.
That emission can last up to minutes; the longest radio train we have observed lasted
6.5\,min, coincident with a CAMS meteor detection with optical magnitude of
$-9$. In general, we find that brighter optical meteors result in
radio trains that remain visible for longer, as shown in
Fig.\,\ref{fig:magnitude}a.

The detected radio emission is integrated over a 2-second time interval. 
Based on the decay of brightness in subsequent 2-s intervals, the short trails 
are mostly emission from the meteor's persistent train, rather than from the meteor itself.

The optical trails of CAMS detected meteors start and end at higher
altitudes for brighter meteors, and while the radio trains show a
similar dependency (Fig.\,\ref{fig:magnitude}b and c), the radio
emission is first detected at lower altitudes
($h=101\pm4$\,km) compared to optical emission ($h=107\pm6$\,km). The
altitude at which the optical and radio emission ends is comparable
for optical ($h=94\pm6$\,km) and radio ($h=92\pm6$\,km).

\begin{figure*}
  \centering
  \scalebox{1.04}{
  \includegraphics[width=0.45\linewidth]{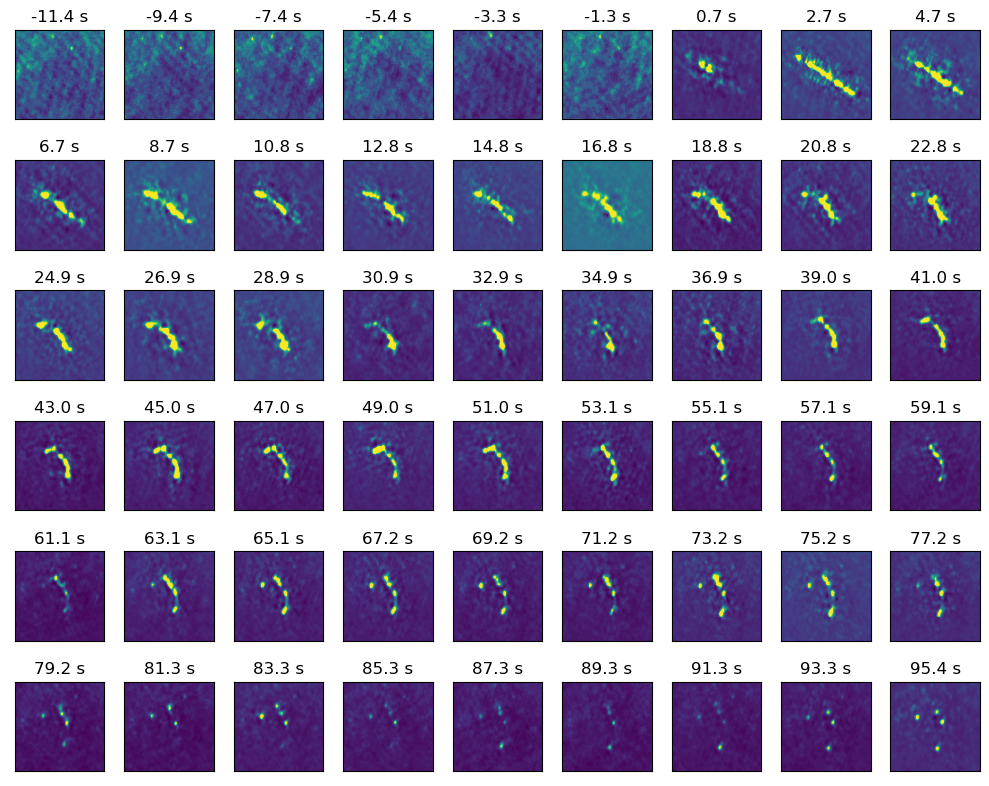}\hspace{5mm}
  \includegraphics[width=0.47\linewidth]{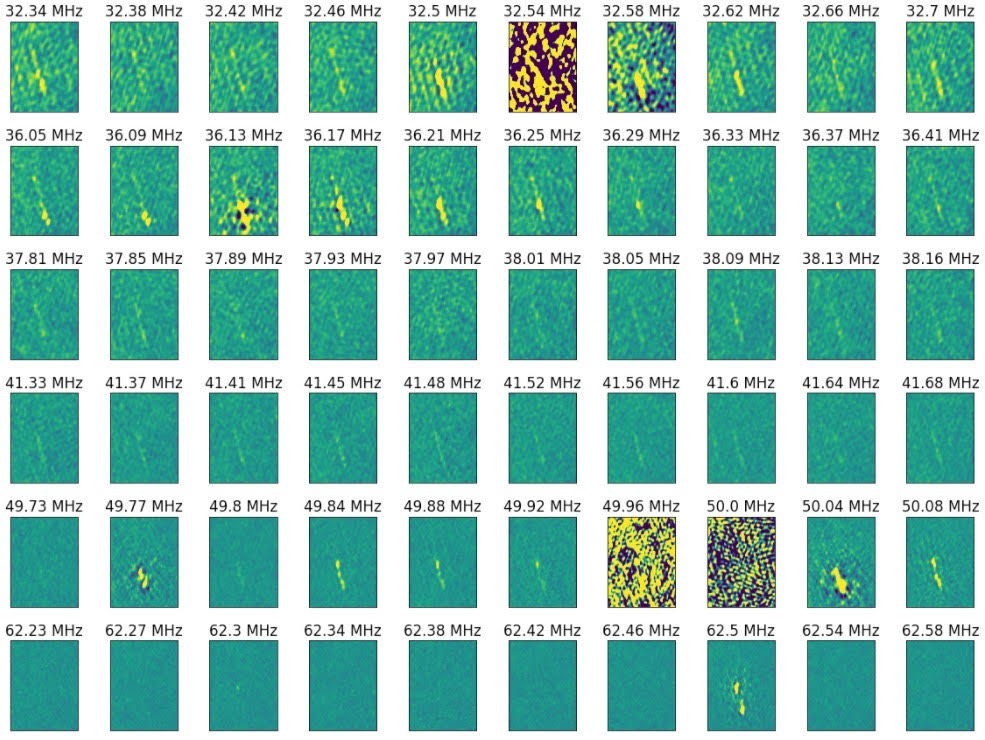}}
  \caption{(Left:) The temporal and spatial evolution of a persistent
    radio meteor train, showing distortion by high altitude winds. The
    time is referenced to the beginning of the optical meteor
    trail. (Right:) Frequency evolution of a meteor train from a
    bright optical meteor. In channels where the emission from the
    meteor was very bright, the calibration has failed. That is the
    case in the bands where narrow-band terrestrial radio emission is
    reflected from the ionization train. Known transmitters operate at
    32.54\,MHz (TV transmitter), 49.97 and 49.99\,MHz (reflection of
    the BRAMS meteor radars; \citealt{2016EGUGA..1811624L}) and
    50.00\,MHz (amateur radio transmissions).}
  \label{fig:time-evolution}
\end{figure*}

Some of the radio trains that were persistent for several minutes were
distorted by high altitude winds, much like optical persistent trains. 
Figure\,\ref{fig:time-evolution} shows an example of the spatial 
evolution of the radio train with time. The peak intensity of the 
radio meteor train decreased over time.

Some frequency channels show reflected narrow-band artificial radio sources, but all
channels contain a broadband radio component that is not likely from
artificial radio sources. For bright radio meteors, the 
signal-to-noise ratio was sufficient to
detect this broadband emission in almost all individual channels for 
frequencies below $\sim50$\,MHz. An example is shown
in Fig.\,\ref{fig:time-evolution}.

Further insight was obtained when repeating the AARTFAAC observations 
during the 2020 Geminid meteor and the 2021 Quadrantid meteor shower. 
Unfortunately, cloudy weather prevented simultaneous CAMS
video observations. During that campaign, the all-sky radio images 
showed significantly fewer radio meteors compared to the observations of the
Perseids.

\section{Discussion}
We can exclude an important contribution from forward scattered
narrow band terrestrial radio emission, which is frequency resolved in
our observations. Broadband emission would require a source of very broad band 
(30--50\,MHz) terrestrial radio emission, which we consider unlikely.

Of the proposed intrinsic emission scenarios, chemically produced 
suprathermal electrons \citep{2020JGRA..12528053O} could cause lingering
radiation over minutes timescale if those electrons are captured by
atoms and molecules that slowly diffuse into the train. Other proposed mechanisms such as
Langmuir waves \citep{2015JGRA..120.9916O}, free-free emission
\citep{2018Ge&Ae..58..693F} and transition radiation
\citep{2020JGRA..12528053O}, would be expected strong in the meteor
itself, but we see the radiation increase in intensity following the
meteor head before fading.

There are several possible reasons why fewer meteor trails were detected during
the Geminids and Quadrantids. Geminids and Quadrantids are known to reach
peak brightness at lower altitudes than the Perseids and show generally weaker
optical persistent trains  \citep{2006MeteorShowers}. In our case, the lower 
entry velocities of the
Geminids and Quadrantids in comparison to the Perseids is perhaps not to blame.
We note that several slower sporadic meteors of similar brightness
were observed in radio during the Perseids meteor shower, see Fig.\,\ref{fig:magnitude}a. 

The atmospheric conditions may be important. During the Perseids, the
summer weather provided hot and dry conditions, while the
Geminids and Quadrantids were observed during the rainy winter season.

Finally, it is possible that the local sidereal time was important. 
The scattering of bright celestial radio sources, as suggested in 
\cite{2015JGRA..120.9916O} was rejected in that paper with an argument involving the
fact the earlier LWA detections are very bright. Since we have detected
much weaker radio meteors, that argument does not hold here.

During the Perseids, the Galactic center and Galactic plane, which is the
source of most low-frequency radio emission, was above the LOFAR
horizon, while this was not the case during the Geminids and
Quadrantids.

\section{Conclusions}
The AARTFAAC observations presented here show that persistent radio emission 
from meteors coincides with their optical
trajectories, and can be detected for meteors of magnitude $\sim0$ and
brighter. Aside from a narrow-band scattering of artificial radio sources, there is also a 
generally broadband emission detected for frequencies below
50\,MHz. The continuum radio emission is first detected at lower altitudes 
compared to the optical emission, while both radio and optical emission end 
at similar altitudes. For the brightest optical meteors,
we find that the radio emission can persist for several minutes. 
Persistent radio trains are affected by high altitude winds like
the optical persistent trains.

Analysis of the AARTFAAC observations is ongoing, so it is currently
not possible to distinguish between the suggested origins of the broadband 
radio emission. 

\vspace{5mm}
\noindent
\textbf{Acknowledgements:} We thank Albert van Duin and Edwin van Dijk
for the use of facilities at the Edgar Getreuer Observatory in
Burlage. We acknowledge interesting discussions with the researchers
in the BRAMS project. We thank all camera operators and other
contributors to CAMS-BeNeLux. PJ acknowledges support from NASA grant
80NSSC19K0563 (Solar System Workings). 

\vspace{5mm}
\noindent
\textbf{Supplemental materials:} Properties of the radio meteors observed
simultaneously with CAMS are available as \citep{zenodo_aartfaac_cams_csv}.

\bibliography{perseids1} 

\begin{thebibliography}{}
\expandafter\ifx\csname natexlab\endcsname\relax\def\natexlab#1{#1}\fi
\providecommand{\url}[1]{\href{#1}{#1}}
\providecommand{\dodoi}[1]{doi:~\href{http://doi.org/#1}{\nolinkurl{#1}}}
\providecommand{\doeprint}[1]{\href{http://ascl.net/#1}{\nolinkurl{http://ascl.net/#1}}}
\providecommand{\doarXiv}[1]{\href{https://arxiv.org/abs/#1}{\nolinkurl{https://arxiv.org/abs/#1}}}

\bibitem[{Dijkema {et~al.}(2021{\natexlab{a}})Dijkema, Bassa, Kuiack,
  Jenniskens, Johannink, Wijers, \& Fallows}]{zenodo_aartfaac_cams_csv}
Dijkema, T.~J., Bassa, C., Kuiack, M., {et~al.} 2021{\natexlab{a}}, {Annotated
  data of simultaneous broadband radio and optical emission of meteor trains
  imaged by LOFAR / AARTFAAC and CAMS},  Zenodo, \dodoi{10.5281/zenodo.5644202}

\bibitem[{Dijkema {et~al.}(2021{\natexlab{b}})Dijkema, Bassa, Kuiack,
  Jenniskens, Wijers, \& Fallows}]{zenodo_mp4_perseids}
---. 2021{\natexlab{b}}, {Time-lapse of AARTFAAC detections of radio meteors in
  Perseids 2020}, 1.0,  Zenodo, \dodoi{10.5281/zenodo.5595288}

\bibitem[{{Filonenko}(2018)}]{2018Ge&Ae..58..693F}
{Filonenko}, A.~D. 2018, Geomagnetism and Aeronomy, 58, 693,
  \dodoi{10.1134/S0016793218050055}

\bibitem[{{Jenniskens}(2006)}]{2006MeteorShowers}
{Jenniskens}, P. 2006, {Meteor Showers and their Parent Comets} (Cambridge
  University Press, Cambridge, UK), 790 pp

\bibitem[{{Jenniskens} {et~al.}(2011){Jenniskens}, {Gural}, {Dynneson},
  {Grigsby}, {Newman}, {Borden}, {Koop}, \& {Holman}}]{2011Icar..216...40J}
{Jenniskens}, P., {Gural}, P.~S., {Dynneson}, L., {et~al.} 2011, Icarus, 216,
  40, \dodoi{10.1016/j.icarus.2011.08.012}

\bibitem[{{Kuiack} {et~al.}(2019){Kuiack}, {Huizinga}, {Molenaar}, {Prasad},
  {Rowlinson}, \& {Wijers}}]{2019MNRAS.482.2502K}
{Kuiack}, M., {Huizinga}, F., {Molenaar}, G., {et~al.} 2019, MNRAS, 482, 2502,
  \dodoi{10.1093/mnras/sty2810}

\bibitem[{{Kuiack} {et~al.}(2021){Kuiack}, {Wijers}, {Shulevski}, {Rowlinson},
  {Huizinga}, {Molenaar}, \& {Prasad}}]{2021MNRAS.505.2966K}
{Kuiack}, M., {Wijers}, R. A.~M.~J., {Shulevski}, A., {et~al.} 2021, MNRAS,
  505, 2966, \dodoi{10.1093/mnras/stab1504}

\bibitem[{{Lamy} {et~al.}(2016){Lamy}, {Ranvier}, {Anciaux}, {Gamby},
  {Calders}, {T{\'e}tard}, \& {De Keyser}}]{2016EGUGA..1811624L}
{Lamy}, H., {Ranvier}, S., {Anciaux}, M., {et~al.} 2016, in EGU General
  Assembly Conference Abstracts, EGU General Assembly Conference Abstracts,
  EPSC2016--11624

\bibitem[{{Lamy} {et~al.}(2011){Lamy}, {Ranvier}, {de Keyser}, {Calders},
  {Gamby}, \& {Verbeeck}}]{2011MSSB}
{Lamy}, H., {Ranvier}, S., {de Keyser}, J., {et~al.} 2011, in Meteoroids: The
  smallest solar system bodies. Proceedings of the Meteoroid Conference held in
  Breckenridge, Co, USA, May 24-28, 2010, ed. W.~J. Cooke, D.~E. Moser, B.~F.
  Hardin, \& D.~Janches, 551--356

\bibitem[{{Obenberger} {et~al.}(2016{\natexlab{a}}){Obenberger}, {Taylor}, \&
  {Holmes}}]{2016AGUFM.P33E..06O}
{Obenberger}, K., {Taylor}, G.~B., \& {Holmes}, J.~M. 2016{\natexlab{a}}, in
  AGU Fall Meeting Abstracts, P33E--06

\bibitem[{{Obenberger} {et~al.}(2016{\natexlab{b}}){Obenberger}, {Dowell},
  {Hancock}, {Holmes}, {Pedersen}, {Schinzel}, \&
  {Taylor}}]{2016JGRA..121.6808O}
{Obenberger}, K.~S., {Dowell}, J.~D., {Hancock}, P.~J., {et~al.}
  2016{\natexlab{b}}, Journal of Geophysical Research (Space Physics), 121,
  6808, \dodoi{10.1002/2016JA022606}

\bibitem[{{Obenberger} {et~al.}(2020){Obenberger}, {Holmes}, {Ard}, {Dowell},
  {Shuman}, {Taylor}, {Varghese}, \& {Viggiano}}]{2020JGRA..12528053O}
{Obenberger}, K.~S., {Holmes}, J.~M., {Ard}, S.~G., {et~al.} 2020, Journal of
  Geophysical Research (Space Physics), 125, e28053,
  \dodoi{10.1029/2020JA028053}

\bibitem[{{Obenberger} {et~al.}(2015){Obenberger}, {Taylor}, {Lin}, {Dowell},
  {Schinzel}, \& {Stovall}}]{2015JGRA..120.9916O}
{Obenberger}, K.~S., {Taylor}, G.~B., {Lin}, C.~S., {et~al.} 2015, Journal of
  Geophysical Research (Space Physics), 120, 9916, \dodoi{10.1002/2015JA021229}

\bibitem[{{Obenberger} {et~al.}(2014){Obenberger}, {Taylor}, {Hartman},
  {Dowell}, {Ellingson}, {Helmboldt}, {Henning}, {Kavic}, {Schinzel},
  {Simonetti}, {Stovall}, \& {Wilson}}]{2014ApJ...788L..26O}
{Obenberger}, K.~S., {Taylor}, G.~B., {Hartman}, J.~M., {et~al.} 2014, ApJ,
  788, L26, \dodoi{10.1088/2041-8205/788/2/L26}

\bibitem[{{Prasad} {et~al.}(2016){Prasad}, {Huizinga}, {Kooistra}, {van der
  Schuur}, {Gunst}, {Romein}, {Kuiack}, {Molenaar}, {Rowlinson}, {Swinbank}, \&
  {Wijers}}]{2016JAI.....541008P}
{Prasad}, P., {Huizinga}, F., {Kooistra}, E., {et~al.} 2016, Journal of
  Astronomical Instrumentation, 5, 1641008, \dodoi{10.1142/S2251171716410087}

\bibitem[{{Prentice} {et~al.}(1947){Prentice}, {Lovell}, \&
  {Banwell}}]{1947MNRAS.107..155P}
{Prentice}, J.~P.~M., {Lovell}, A.~C.~B., \& {Banwell}, C.~J. 1947, MNRAS, 107,
  155, \dodoi{10.1093/mnras/107.2.155}

\bibitem[{{Price} \& {Blum}(2000)}]{2000EMP82-83}
{Price}, C., \& {Blum}, M. 2000, Earth, Moon Planets, 82-83, 545–554

\bibitem[{{Roggemans}(2020)}]{2020eMetN...5..400R}
{Roggemans}, P. 2020, eMeteorNews, 5, 400

\bibitem[{{Shulevski} {et~al.}(2021){Shulevski}, {Franzen}, {Williams},
  {Vernstrom}, {Gehlot}, {Kuiack}, \& {Wijers}}]{2021arXiv210315160S}
{Shulevski}, A., {Franzen}, T.~M.~O., {Williams}, W.~L., {et~al.} 2021, arXiv
  e-prints, arXiv:2103.15160.
\newblock \doarXiv{2103.15160}

\bibitem[{{van Haarlem} {et~al.}(2013){van Haarlem}, {Wise}, {Gunst}, {Heald},
  {McKean}, {Hessels}, {de Bruyn}, {Nijboer}, {Swinbank}, {Fallows},
  {Brentjens}, {Nelles}, {Beck}, {Falcke}, {Fender}, {H{\"o}randel},
  {Koopmans}, {Mann}, {Miley}, {R{\"o}ttgering}, {Stappers}, {Wijers},
  {Zaroubi}, {van den Akker}, {Alexov}, {Anderson}, {Anderson}, {van Ardenne},
  {Arts}, {Asgekar}, {Avruch}, {Batejat}, {B{\"a}hren}, {Bell}, {Bell}, {van
  Bemmel}, {Bennema}, {Bentum}, {Bernardi}, {Best}, {B{\^\i}rzan}, {Bonafede},
  {Boonstra}, {Braun}, {Bregman}, {Breitling}, {van de Brink}, {Broderick},
  {Broekema}, {Brouw}, {Br{\"u}ggen}, {Butcher}, {van Cappellen}, {Ciardi},
  {Coenen}, {Conway}, {Coolen}, {Corstanje}, {Damstra}, {Davies}, {Deller},
  {Dettmar}, {van Diepen}, {Dijkstra}, {Donker}, {Doorduin}, {Dromer}, {Drost},
  {van Duin}, {Eisl{\"o}ffel}, {van Enst}, {Ferrari}, {Frieswijk}, {Gankema},
  {Garrett}, {de Gasperin}, {Gerbers}, {de Geus}, {Grie{\ss}meier}, {Grit},
  {Gruppen}, {Hamaker}, {Hassall}, {Hoeft}, {Holties}, {Horneffer}, {van der
  Horst}, {van Houwelingen}, {Huijgen}, {Iacobelli}, {Intema}, {Jackson},
  {Jelic}, {de Jong}, {Juette}, {Kant}, {Karastergiou}, {Koers}, {Kollen},
  {Kondratiev}, {Kooistra}, {Koopman}, {Koster}, {Kuniyoshi}, {Kramer},
  {Kuper}, {Lambropoulos}, {Law}, {van Leeuwen}, {Lemaitre}, {Loose}, {Maat},
  {Macario}, {Markoff}, {Masters}, {McFadden}, {McKay-Bukowski}, {Meijering},
  {Meulman}, {Mevius}, {Middelberg}, {Millenaar}, {Miller-Jones}, {Mohan},
  {Mol}, {Morawietz}, {Morganti}, {Mulcahy}, {Mulder}, {Munk}, {Nieuwenhuis},
  {van Nieuwpoort}, {Noordam}, {Norden}, {Noutsos}, {Offringa}, {Olofsson},
  {Omar}, {Orr{\'u}}, {Overeem}, {Paas}, {Pandey-Pommier}, {Pandey}, {Pizzo},
  {Polatidis}, {Rafferty}, {Rawlings}, {Reich}, {de Reijer}, {Reitsma},
  {Renting}, {Riemers}, {Rol}, {Romein}, {Roosjen}, {Ruiter}, {Scaife}, {van
  der Schaaf}, {Scheers}, {Schellart}, {Schoenmakers}, {Schoonderbeek},
  {Serylak}, {Shulevski}, {Sluman}, {Smirnov}, {Sobey}, {Spreeuw}, {Steinmetz},
  {Sterks}, {Stiepel}, {Stuurwold}, {Tagger}, {Tang}, {Tasse}, {Thomas},
  {Thoudam}, {Toribio}, {van der Tol}, {Usov}, {van Veelen}, {van der Veen},
  {ter Veen}, {Verbiest}, {Vermeulen}, {Vermaas}, {Vocks}, {Vogt}, {de Vos},
  {van der Wal}, {van Weeren}, {Weggemans}, {Weltevrede}, {White}, {Wijnholds},
  {Wilhelmsson}, {Wucknitz}, {Yatawatta}, {Zarka}, {Zensus}, \& {van
  Zwieten}}]{2013A&A...556A...2V}
{van Haarlem}, M.~P., {Wise}, M.~W., {Gunst}, A.~W., {et~al.} 2013, A\&A, 556,
  A2, \dodoi{10.1051/0004-6361/201220873}

\bibitem[{{Zhang} {et~al.}(2018){Zhang}, {Hancock}, {Devillepoix}, {Wayth},
  {Beardsley}, {Crosse}, {Emrich}, {Franzen}, {Gaensler}, {Horsley},
  {Johnston-Hollitt}, {Kaplan}, {Kenney}, {Morales}, {Pallot}, {Steele},
  {Tingay}, {Trott}, {Walker}, {Williams}, {Wu}, {Ji}, \&
  {Ma}}]{2018MNRAS.477.5167Z}
{Zhang}, X., {Hancock}, P., {Devillepoix}, H.~A.~R., {et~al.} 2018, Monthly
  Notices of the RAS, 477, 5167, \dodoi{10.1093/mnras/sty930}

\end{thebibliography}
\bibliographystyle{aasjournal}

\end{document}